# Accelerators for Medical Applications—Radio Frequency Powering

*E. Montesinos*
CERN, Geneva, Switzerland

**Abstract**
This paper reviews the main types of radio-frequency powering systems which may be used for medical applications. It gives the essentials on vacuum tubes, including tetrodes, klystrons, and inductive output tubes, and the essentials on transistors. The basics of combining systems, splitting systems, and transmission lines are discussed. The paper concludes with a case study specific to medical applications, including overall efficiency and cost analysis regarding the various available technologies.

**Keywords**
Vacuum tube; tetrode; klystron; IOT; transistor; transmission lines.

## 1 Introduction

Cost is a very important factor for all projects, and this is particularly true for medical applications. Figure 1 summarizes, at a glance, the ratios that are currently applied when talking about radio frequency (RF)-power systems.

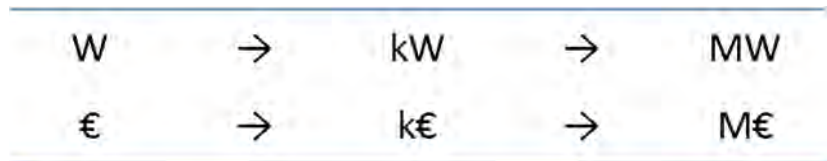

**Fig. 1:** Relationship between Watts and Euros in RF-power systems

We will describe, in this document, the technologies used to build RF-power amplifiers, and the high technicity involved explains such high numbers.

## 2 RF-power basics

When we talk about an RF-power system, one should understand the system amplifying a small RF signal from the generator, in the order of mW, up to the W, kW, or MW level at the Device Under Test input, that could be an accelerating cavity, or any other RF load, as described in Fig. 2.

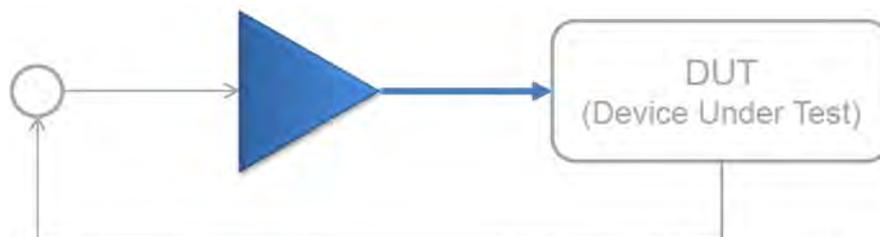

**Fig. 2:** A very simple representation of an RF-power system. It includes the RF-power amplifiers, the transmission lines and the Fundamental Power Coupler.

Some specific parameters characterize RF-power systems, such as wavelength, frequency, and Decibel (dB). We will describe these basic concepts in the following paragraphs.

## 2.1 Wavelength and frequency

The wavelength (Fig. 3) and the frequency are linked with the following formulas:

$$\lambda = \frac{c}{f\sqrt{\varepsilon}} \qquad (1)$$

$$f = \frac{c}{\lambda\sqrt{\varepsilon}} \qquad (2)$$

where

$\lambda$ = wavelength in meters (m),

$c$ = velocity of light (m/s)—(~300 000 000 m/s),

$f$ = frequency in hertz (Hz), and

$\varepsilon$ = dielectric constant of the propagation medium.

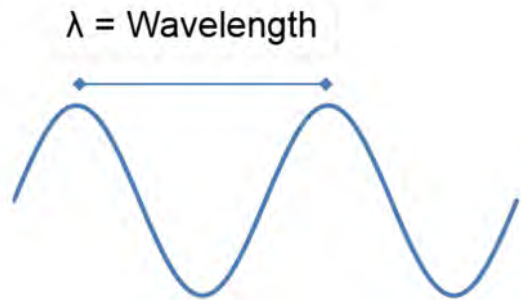

**Fig. 3:** The wavelength is the distance between two points reaching exactly the same potential of a sine wave oscillating at a given frequency.

One can see that the medium is important as the wavelength will be directly proportional to its propagation dielectric constant. In air or vacuum, $\varepsilon$ is equal to 1, if the medium is PTFE, as in our coaxial cable from the antenna to the television, $\varepsilon$ is around 2.2. The wavelength will then be shortened by $\sqrt{2.2}$. One will have to keep this in mind when designing an RF-power system.

## 2.2 Electromagnetic spectrum and radiofrequency spectrum

The RF spectrum is a fraction of the electromagnetic (EM) spectrum. It has been defined as starting at 30 kHz and ending at 300 GHz. Respectively, the wavelength is of 10 km reducing to 1 mm. Figure 4 shows the entire EM spectrum with some examples of current known applications, and Fig. 5 shows the RF spectrum with the uses in our current lives.

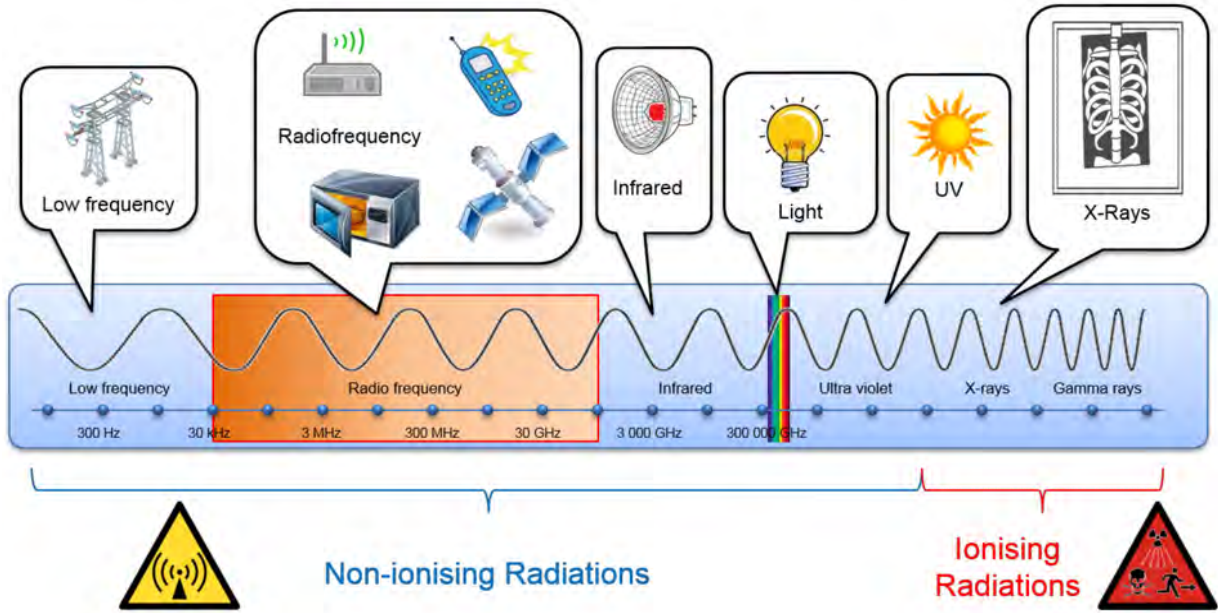

**Fig. 4:** The EM spectrum with respect to frequency range

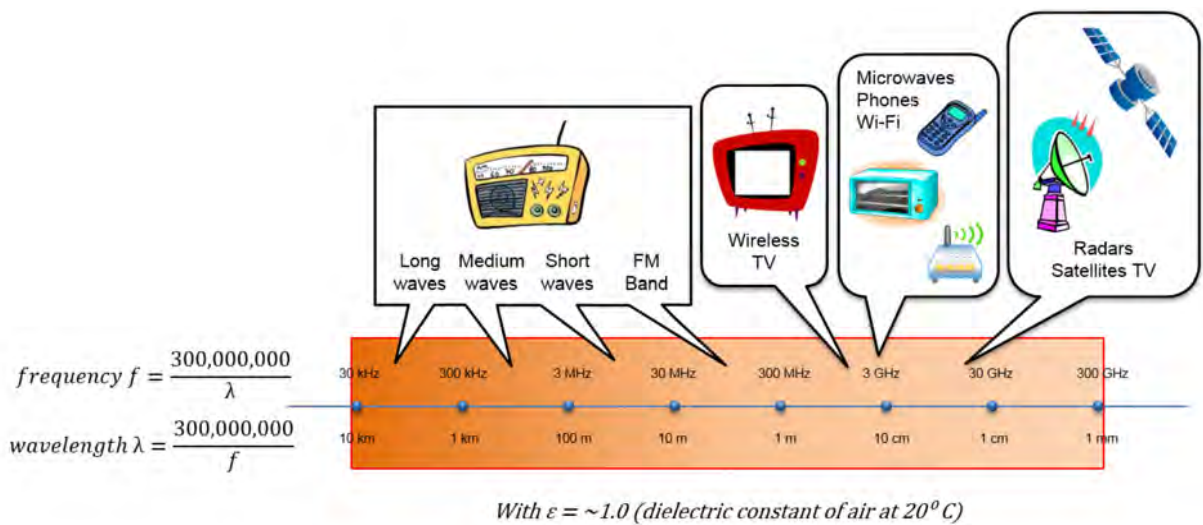

**Fig. 5:** The RF spectrum with relationship between frequency and wavelength

## 2.3 Decibel

A useful unit used with RF-power equipment is the Decibel. This unit allows us to sum the gain or subtract the attenuation instead of multiplying the gain and dividing the attenuation. It makes calculations of power systems easier. The dBm (3) is to quantify absolute values, and the dB (4) to quantify ratio. The definitions of the dB, commonly used in RF power, are:

$$\mathrm{dBm} = 10\,\log_{10}\left(\mathrm{P}_{mW}\right), \tag{3}$$

$$\mathrm{dB} = 10\,\log_{10}\left(P_1 / P_2\right). \tag{4}$$

Several other definitions of the dB exist:

$$\mathrm{dB} = 20\log_{10}\left(V_1 / V_2\right), \tag{5}$$

$$\mathrm{dB}V = 20 \log_{10}\left(V_{Vrms}\right), \tag{6}$$

$$\mathrm{dB}\mu V = 20 \log_{10}\left(V_{\mu V_{rms}}\right), \tag{7}$$

$$\mathrm{dB}c = 10 \log_{10}\left(\frac{P_{carrier}}{P_{signal}}\right). \tag{8}$$

Some absolute values are useful to memorize (Fig. 6). One can switch from dBm to watt applying the following formula (9):

$$x_{dBm} = 10 \, log_{10}\left(P_{mW}\right) \leftrightarrow P_{mW} = 10\left(x_{dBm}/10\right). \tag{9}$$

| 0 dBm | = | 1 mW |
|---|---|---|
| 30 dBm | = | 1 W |
| 60 dBm | = | 1 kW |
| 90 dBm | = | 1 MW |

**Fig. 6:** Some known dBm to watt values

One can also switch from dB to ratio in power (Fig. 7) following formula (10):

$$x_{dB} = 10 \, log_{10}\left(P/P_{ref}\right) \leftrightarrow P/P_{ref} = 10\left(x_{dB}/10\right). \tag{10}$$

| x (dB) | P/P$_{ref}$ | |
|---|---|---|
| + 0.1 | 1.023 | + 2.5% |
| + 0.5 | 1.122 | + 12% |
| + 1 | 1.259 | + 25% |
| + 3 | 1.995 | 2 |
| - 0.1 | 0.977 | - 2.5% |
| - 0.5 | 0.891 | - 11% |
| - 1 | 0.794 | - 20% |
| - 3 | 0.501 | 0.5 |

**Fig. 7:** Some known ratio values

## 3 RF-power amplifiers

RF-power amplifiers can be sorted into two main families: the vacuum tubes and the transistors. The vacuum tubes consist of three main families: the grid tubes, the linear beam tubes, and the crossed-field tubes. The transistor amplifiers are also named solid state amplifiers (SSA) or solid state power amplifiers (SSPA). In this document, we will look in further detail at the tetrodes, the klystrons, the inductive output tube (IOT), and the laterally diffuse metal oxide semiconductor (LDMOS). Figure 8 shows the different families.

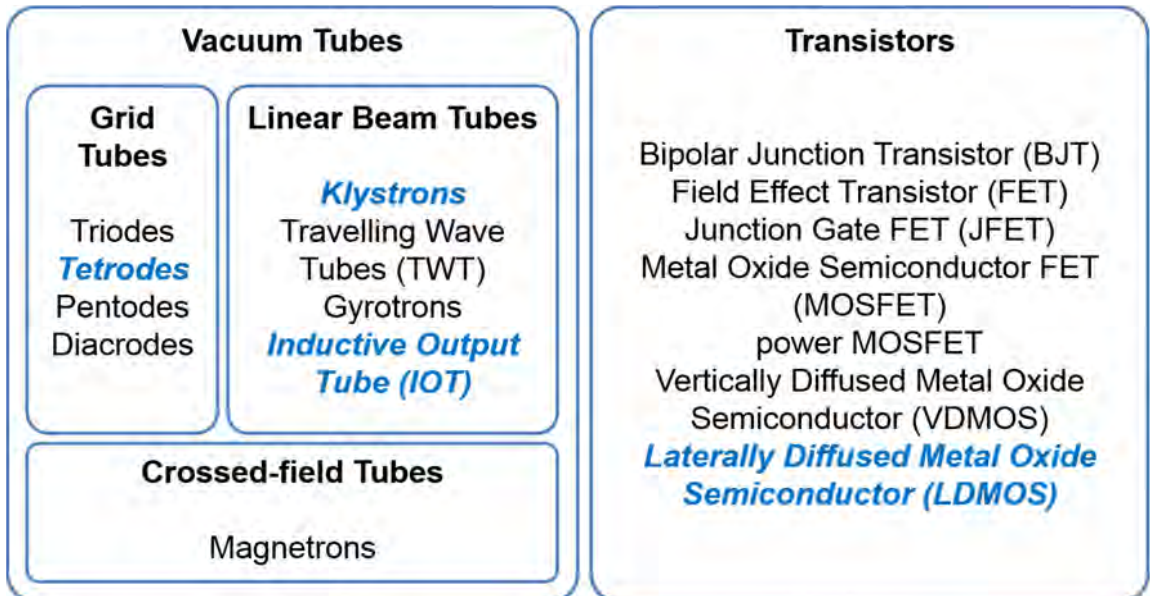

**Fig. 8:** Main list, non-exhaustive, of RF-power-amplifier families

### 3.1 Grid tubes

The grid tube story started more than a century ago in 1904 with the very first diode [1]. Hereunder, the list of the main milestones of the grid tube story. It is very interesting to note that most of the discoveries were made within the first quarter of the last century, even though, almost a century later in 1994, thanks to new fabrication methods, new tubes are still being developed.

1904 Diode, John Ambrose Fleming (Fig. 9) [1];

1906 Audion (first triode), Lee de Forest [2];

1912 Triode as amplifier, Fritz Lowenstein [3];

1913 Triode 'higher vacuum', Harold Arnold [4];

1915 first transcontinental telephone line, Bell [5];

1916 Tetrode, Walter Schottky [6];

1926 Pentode, Bernardus Tellegen [7];

1994 Diacrode, Thales Electron Devices [8].

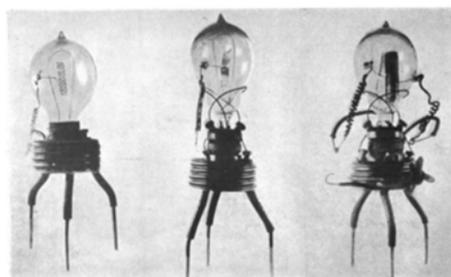

**Fig. 9:** Very first diode invented by John Ambrose Fleming in 1904

Vacuum tube history started with the diode [1]. Looking at Fig. 10, we can identify the heater and the cathode. In this illustration, we are looking at a heated cathode circuit. There is a separate heater and a cathode. There also exists a circuit with a direct-heated cathode. In that case, the cathode also includes the heater. The cathode system is a complex system composed of coated metal, doped with carbides, borides, and other specific components developed by the tube suppliers to ensure good electron emission. Once the cathode is heated, a thermionic emission starts and an electron cloud is generated around the cathode. If we now apply a high voltage on the anode side, these electrons will fly from the cathode to the anode. If we reverse the anode voltage to a negative value, the electrons will remain at the cathode level. We have here a diode.

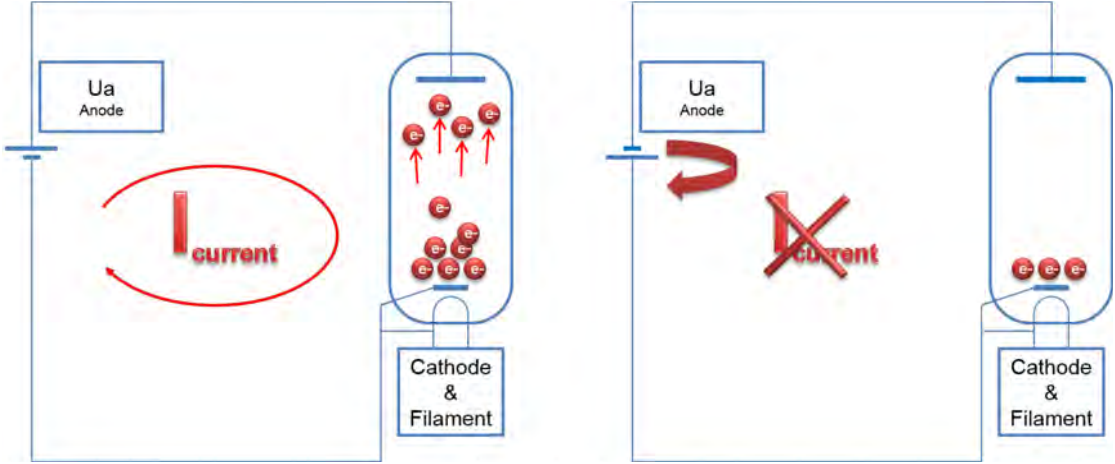

**Fig. 10:** On the left, with a positive voltage on the anode, electrons fly from the grid to the anode. On the right, with a negative voltage on the anode, electrons remain at the grid. This is the basic principle of the very first diode invented by John Ambrose Fleming in 1904 [1].

A few years later, in 1906, Lee de Forest [2] added a control grid in between the cathode and the anode, as showed in Fig. 11. By modulating the voltage applied to the grid, we proportionally modulate the anode current. This is the trans-conductance effect: voltage modulation at the grid is transformed into current modulation at the anode. Indeed, when the grid voltage is less negative than the cathode voltage, the electrons fly to the anode, and when the grid voltage is more negative than the cathode voltage, the electrons remain in the cathode. Unfortunately, there are some limitations in this system. The parasitic capacitor between the grid and the anode gives the system a tendency to oscillate.

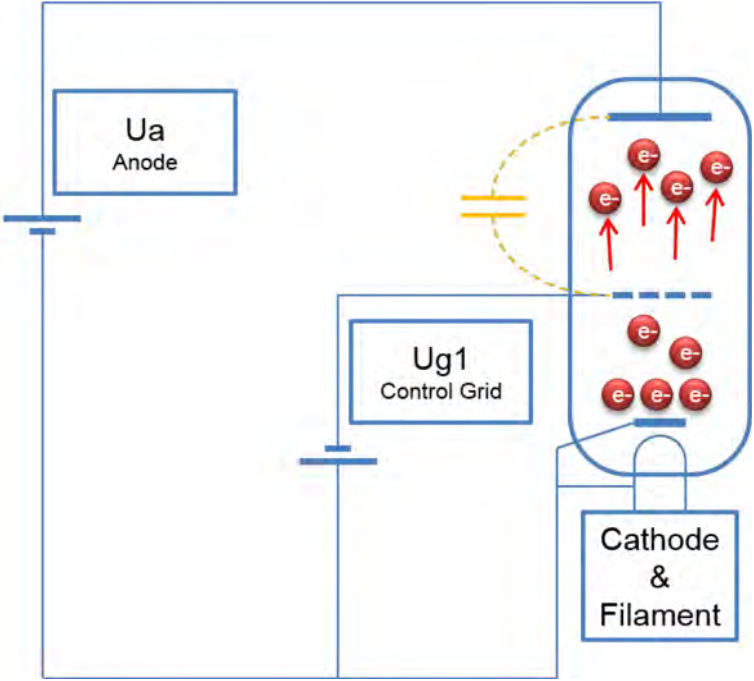

**Fig. 11:** A control grid is inserted in-between the cathode and the anode in order to modulate the electron flux. This is the basic principle of the triode invented by Lee de Forest in 1906 [2].

In order to suppress this tendency to oscillate, a second grid, the screen grid, has been added in between the control grid and the anode. With its positive voltage, lower than the anode, it provides two main advantages. It allows the parasitic capacitor between the control grid and the anode to be decoupled. It also provides better attraction of the electron, as it is close to the control grid and the cathode. This provides a better gain compared to the triode. Unfortunately, this also generates some additional limitations. As sketched in Fig. 12, some of the electrons are accelerated too much and once they hit the anode, they generate secondary electrons which fly from the anode to the control grid. To prevent this effect, tube manufacturers have developed a special treatment of the anode to reduce this secondary emission. Figures 13 and 14 show the CERN SPS based on RS2004 tetrode amplifiers.

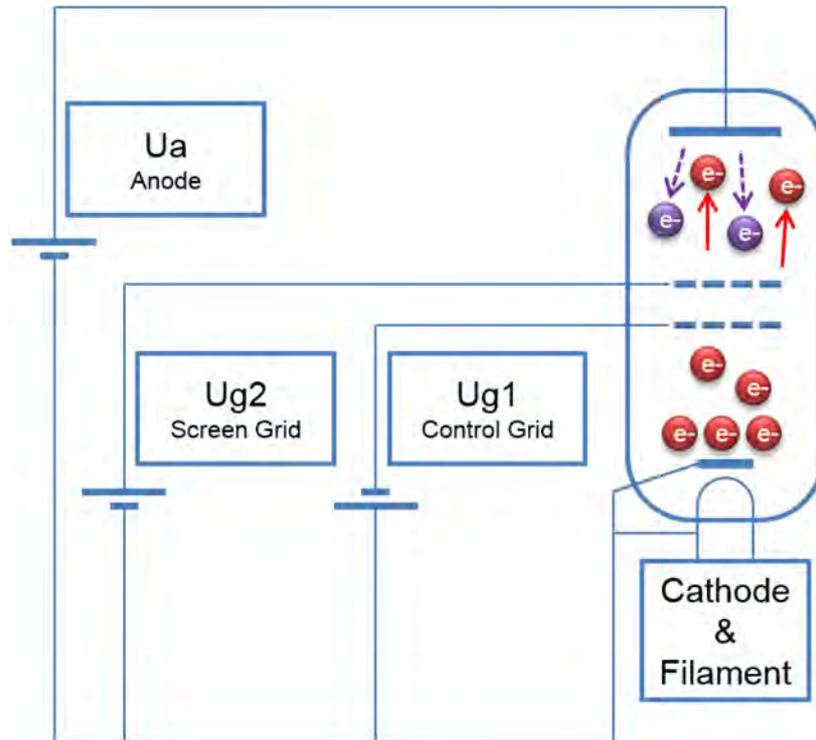

**Fig. 12:** A second grid, the screen grid, is inserted in between the control grid and the anode. This is the basic principle of the tetrode invented by Walter Schottky in 1916 [6].

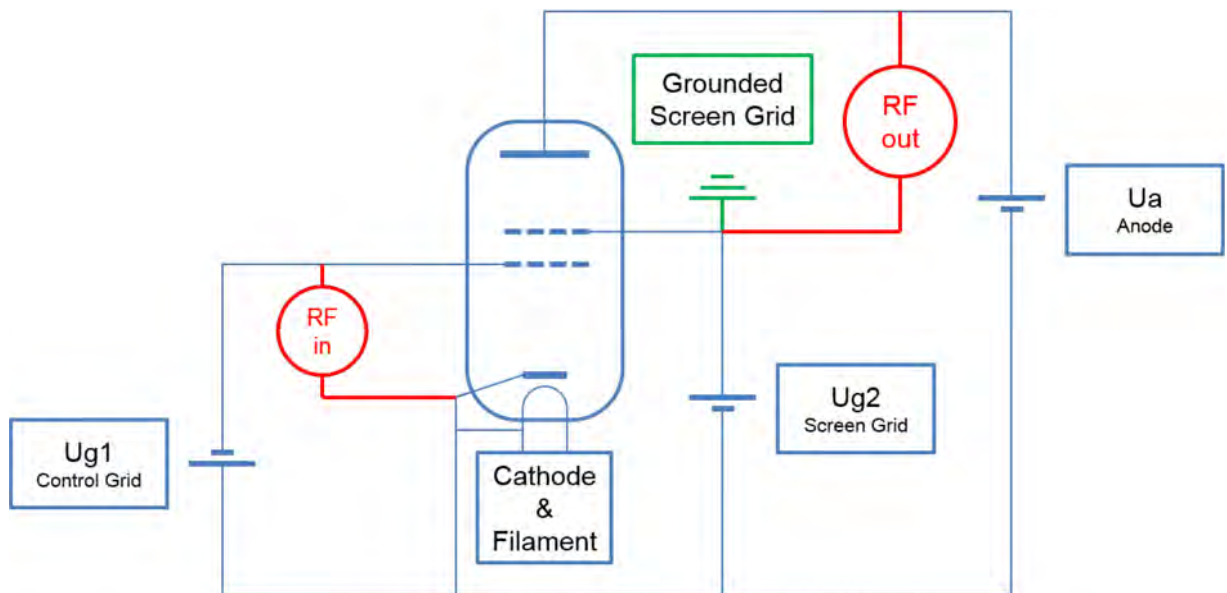

**Fig. 13:** CERN SPS, RS 2004 Tetrode (very) simplified bloc diagram

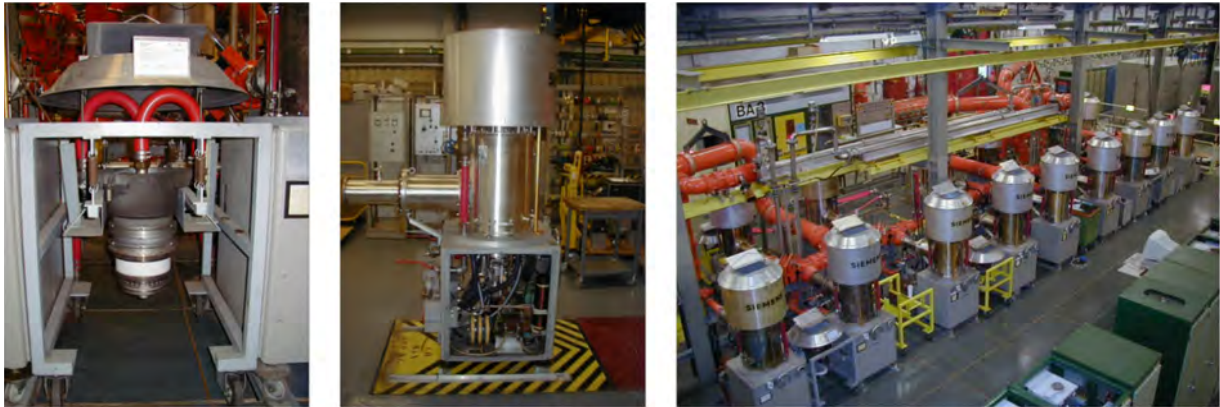

**Fig. 14:** CERN SPS, RS 2004 Tetrode, on the left a trolley (single amplifier), in the centre a transmitter (combination of four amplifiers) and on the right two transmitters (combination of eight amplifiers) delivering 2 x 1 MW @ 200 MHz, into operation since 1976.

An additional grid can be inserted. We then have a pentode. However, the construction complexity of such a tube limited its usage to lower-power systems.

More recently, technical fabrication improvements have been made allowing Thales to construct a Diacrode© [8]. This tube is equivalent to a double-ended tetrode, allowing even more power with a single tube.

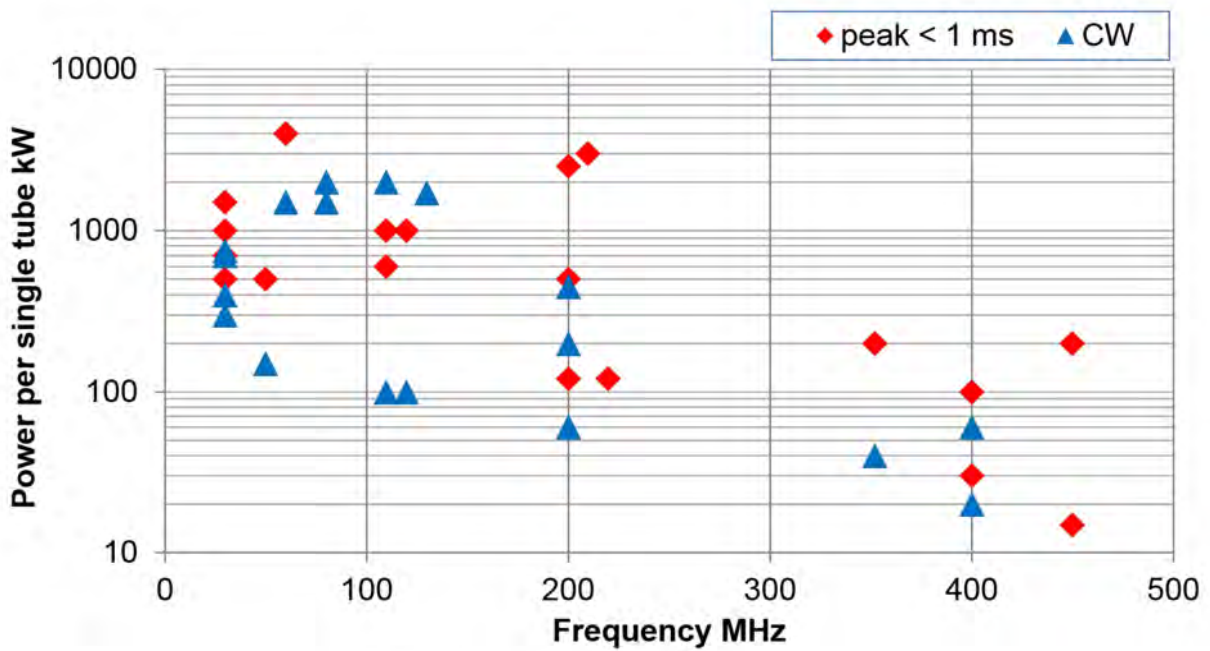

**Fig. 15:** Tetrodes and Diacrodes available from industry

Figure 15 summarizes all tubes currently available from worldwide suppliers. We note that the maximum power is over 1 MW at low frequency, power decreases with frequency, and the frequency range is from a few MHz to 400 MHz.

### 3.2 Linear beam tubes

The linear beam tube story started later, in 1937, with the very first klystron. Hereunder, the list of the main milestones of the linear beam tube story. It is very interesting to note that most of the discoveries have been made within a decade, from 1937 to 1948. Later, as for the grid tubes, thanks to the new fabrication methods, new tubes have been and are still developed:

- 1937 Klystron, Russell and Sigurd Varian [9];
- 1938 IOT, Andrew V. Haeff [10];
- 1939 Reflex klystron, Robert Sutton;
- 1940 Few commercial IOT;
- 1941 Magnetron, Randall and Boot [11];
- 1945 Helix travelling wave tube (TWT), Kompfner [12];
- 1948 Multi MW klystron;
- 1959 Gyrotron, Twiss and Schneider;
- 1963 Multi beam klystron, Zusmanovsky and Korolyov [13];
- 1980 High efficiency IOT.

### 3.2.1 *Klystron*

The klystron is built around a different concept than the grid tube. It uses the velocity modulation of an electron beam to generate high-power RF. The principle, here, is to convert the kinetic energy of the electrons into RF power. Looking at Fig. 16, we can identify the electron gun. It is composed of a thermionic cathode and an anode. We then have a drift space and, at the end, a collector. When applying all the voltages, an electron beam is generated and electrons fly with a constant speed from the gun to the collector through the drift space.

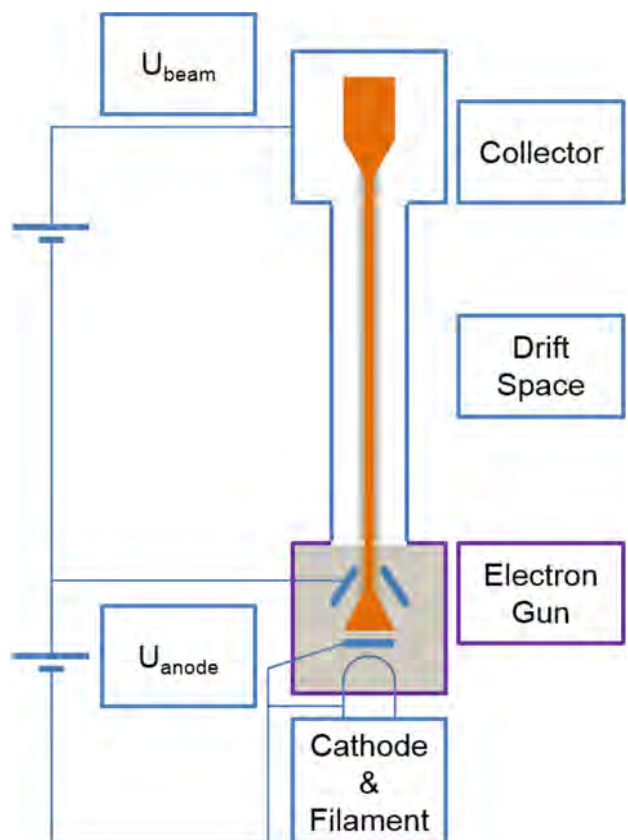

**Fig. 16:** The DC sketch of a klystron

In order to convert this constant electron flux into an RF-power generator, we add cavity resonators. The RF input cavity is the Buncher and the RF output cavity is the Catcher (Fig. 17).

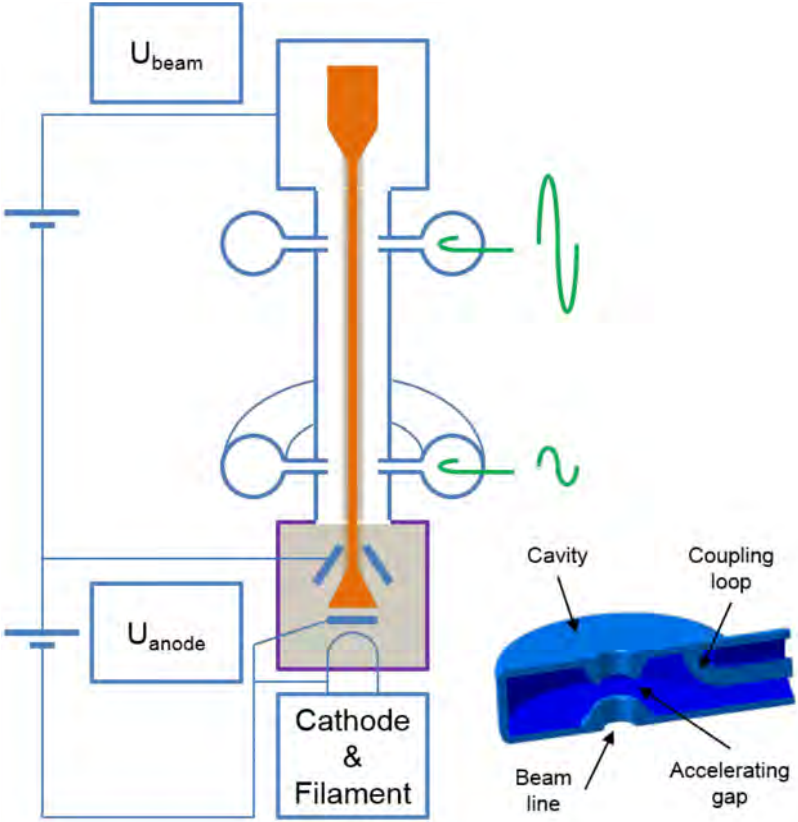

**Fig. 17:** The RF sketch of a klystron. The input cavity is the Buncher. The output cavity is the Catcher

The principle is the following. By applying RF on the Buncher, we modulate the speed of the electrons. Some electrons are accelerated, some are neutral, and some are decelerated. Figure 18 illustrates how we obtain the bunching of the electrons.

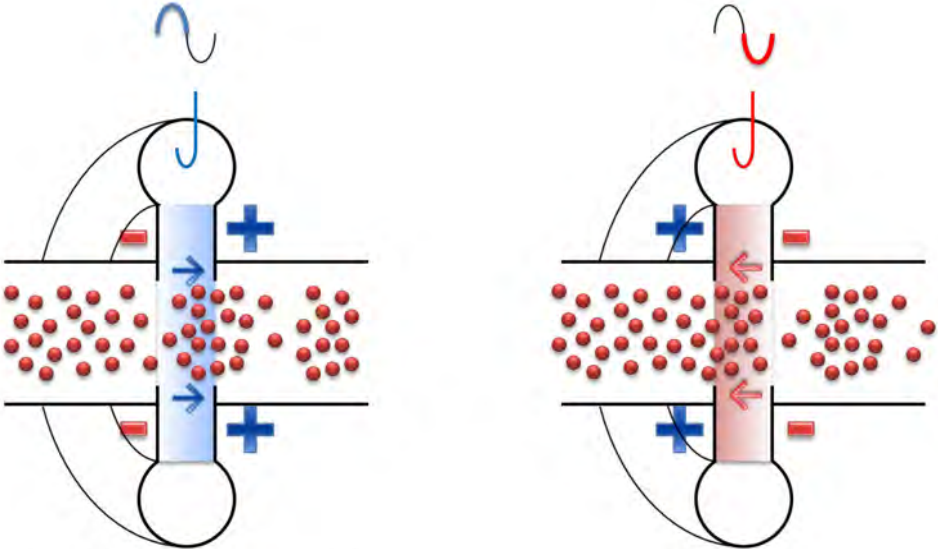

**Fig. 18:** Bunching of the electrons. On the left, when the voltage seen by the electrons at the Buncher cavity gap is positive, electrons are accelerated. On the right, when the voltage seen by the electrons at the Buncher cavity gap is negative, electrons are decelerated.

At the end of the electrons' journey, the Catcher cavity resonates at the same frequency as the input cavity. It is designed to be at the exact position with the maximum number of electrons. The kinetic energy of all these electrons is then converted into voltage and extracted from the output cavity. We then have an RF-power amplifier as shown in Fig. 19. Figure 20 shows the drift space distance between the Buncher cavity and Catcher cavity and how they must be spaced in order to maximize the efficiency of the tube.

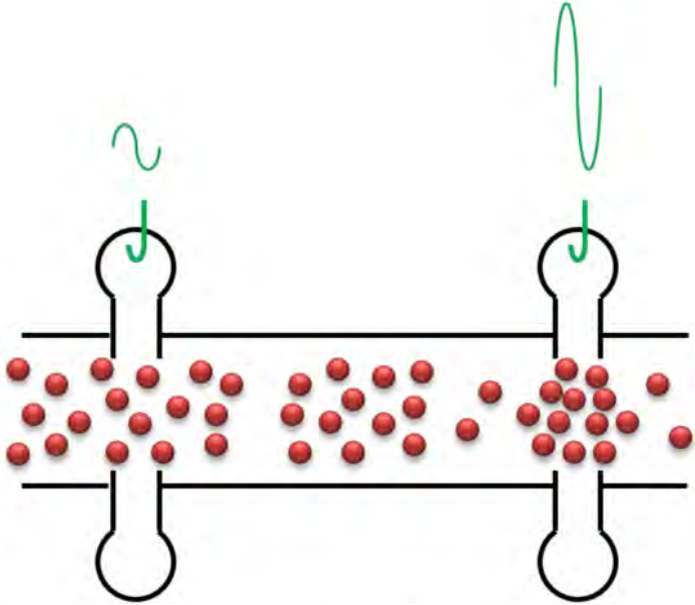

**Fig. 19:** Buncher and Catcher cavities. A constant electron flux before the Buncher cavity is transformed into bunched electrons at the Catcher cavity. The kinetic energy of these bunched electrons is converted into voltage and extracted from the Catcher cavity.

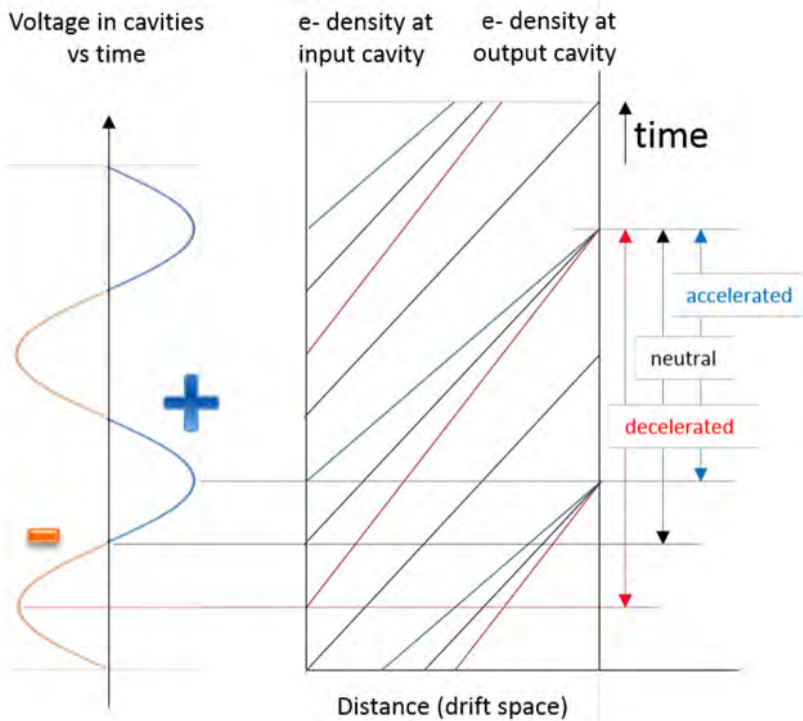

**Fig. 20:** Bunching of electron beam in a klystron. Distance of the drift space allows for maximum electron density at the Catcher cavity plan.

In order to increase the gain of this two-cavity klystron, additional cavities, resonating with the pre-bunched electron beam, are added. These additional cavities generate additional accelerating and decelerating fields. They provide a better bunching, and it is commonly acknowledged that they provide around 10 dB gain per additional cavity.

In order to keep the beam correctly focused in the drift space, focusing magnets are mandatory. They ensure that the electron beam is maintained, as expected and where expected. Figures 21 and 22 show the CERN LHC based on TH2167 klystron amplifiers.

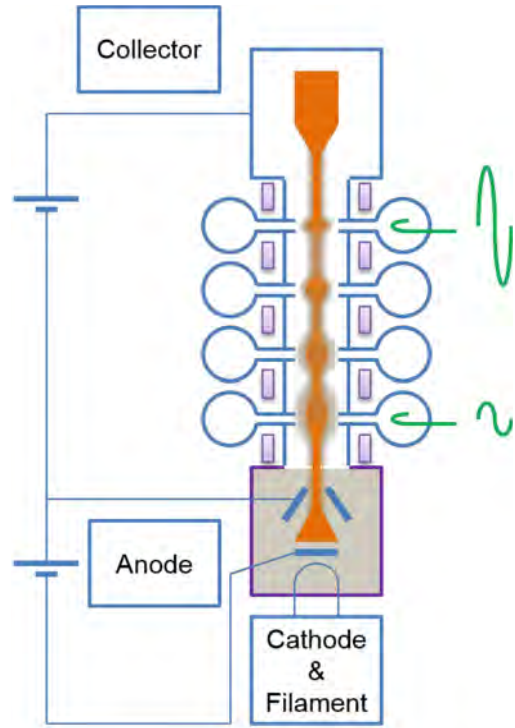

**Fig. 21:** Sketch of a klystron with four cavities and its focusing magnets. The gain of such a device would be around 40 dB.

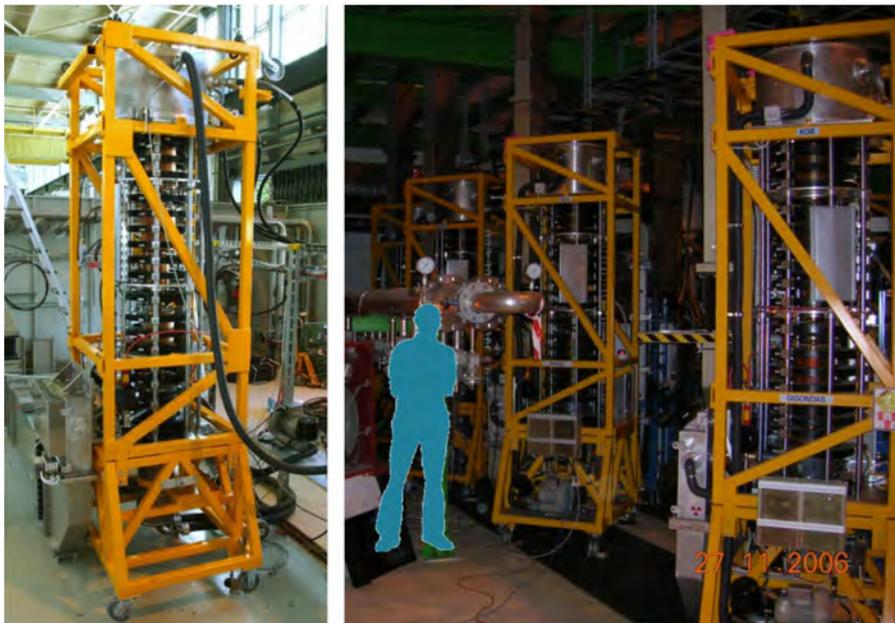

**Fig. 22:** CERN LHC, TH 2167 klystron. On the left, in lab, and on the right in UX45 LHC cavern, 16 klystrons delivering 330 kW @ 400 MHz, in operation since 2008.

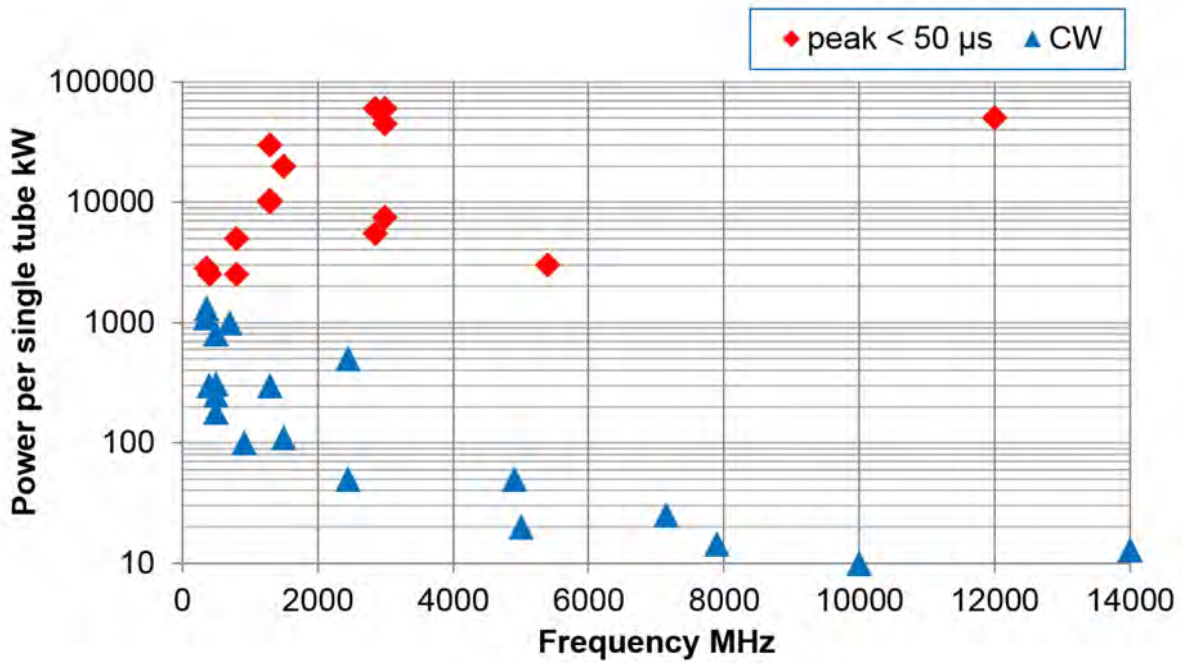

**Fig. 23:** Klystrons available from industry

Figure 23 summarizes all klystrons currently available from worldwide suppliers. We note that the maximum peak power is over 10 MW at low frequency. Continuous wave (CW) power decreases with frequency, and the frequency range is from a few MHz to several GHz.

### 3.2.2 *Inductive output tube*

The IOT is a mix between a triode and a klystron. Here, the principle is to modulate the density of an electron beam with a triode input. We recognize the thermionic cathode and the control grid that modulates the electron emission. On the output side, we have a simplified klystron circuit. We recognize the anode that accelerates the electron beam. Then, we have a short drift space, the Catcher cavity, and the collector. We also have a magnet to keep the beam as expected. Even though the IOT was invented at approximately the same time as the klystron, they were not used before the 1990s. Indeed, IOT gain is lower than the klystron, being approximately in the order of 23 dB, much lower than a five-cavity klystron that will have approximately 50 dB gain. In addition, it also requires a high-voltage power supply, as for the klystron, of around 30 kV to 50 kV, much higher than the 10 kV to 15 kV needed with a tetrode. For all these reasons, it was considered for a long time that IOTs were the sum of all the disadvantages of the tetrodes and of the klystrons. However, with the recent improvement in solid state amplifiers allowing a higher-power driver without tubes, they recently turned out to be an elegant solution where a single RF medium power source is needed. Since the beginning of the new century, they have been implemented in several laboratories around the world. Figure 25 shows the CERN SPS based on TH795 IOT amplifiers.

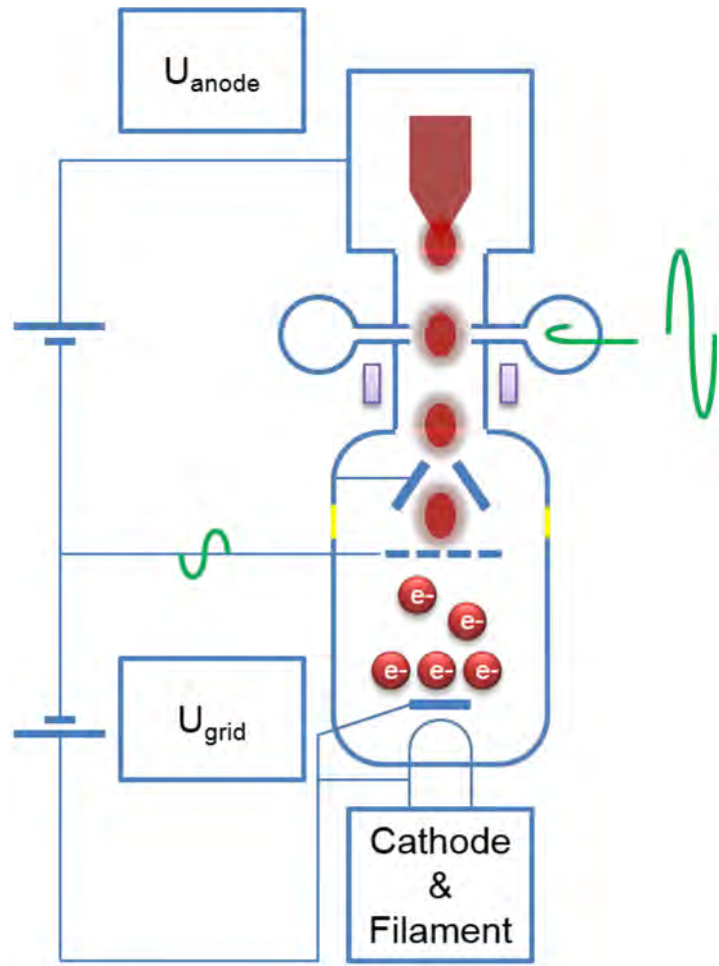

**Fig. 24:** Sketch of an IOT

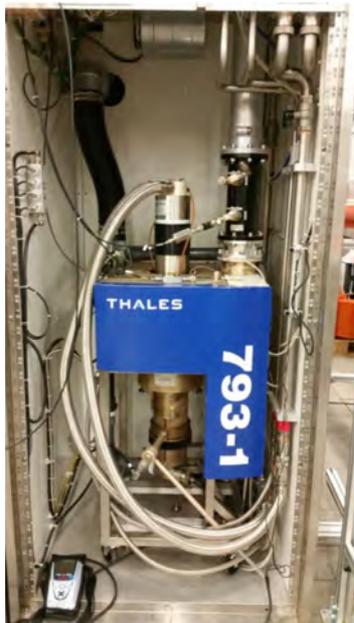
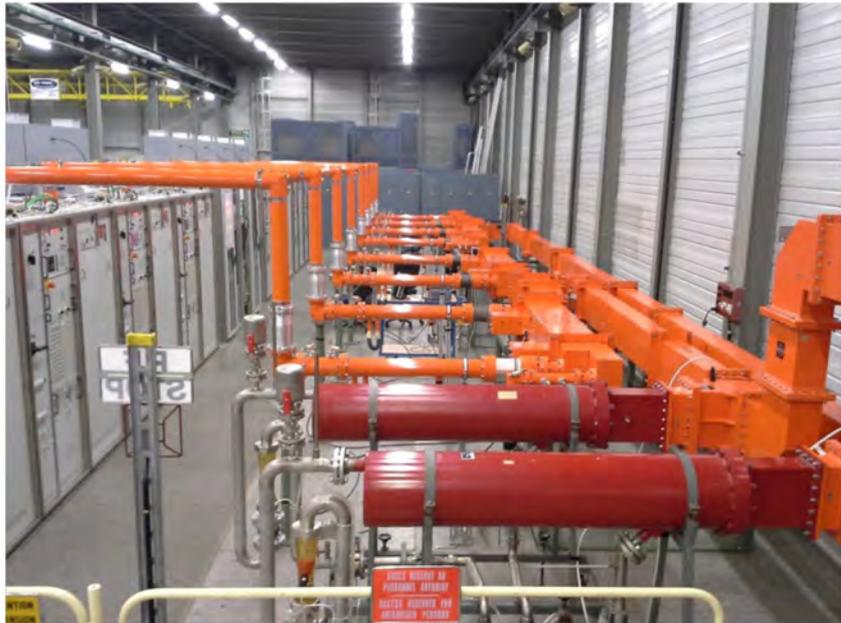

**Fig. 25:** CERN SPS, TH 795 IOT, trolley (single amplifier), and transmitter (combination of amplifiers). Two transmitters of four tubes delivering 2 x 240 kW @ 801 MHz, in operation since 2014.

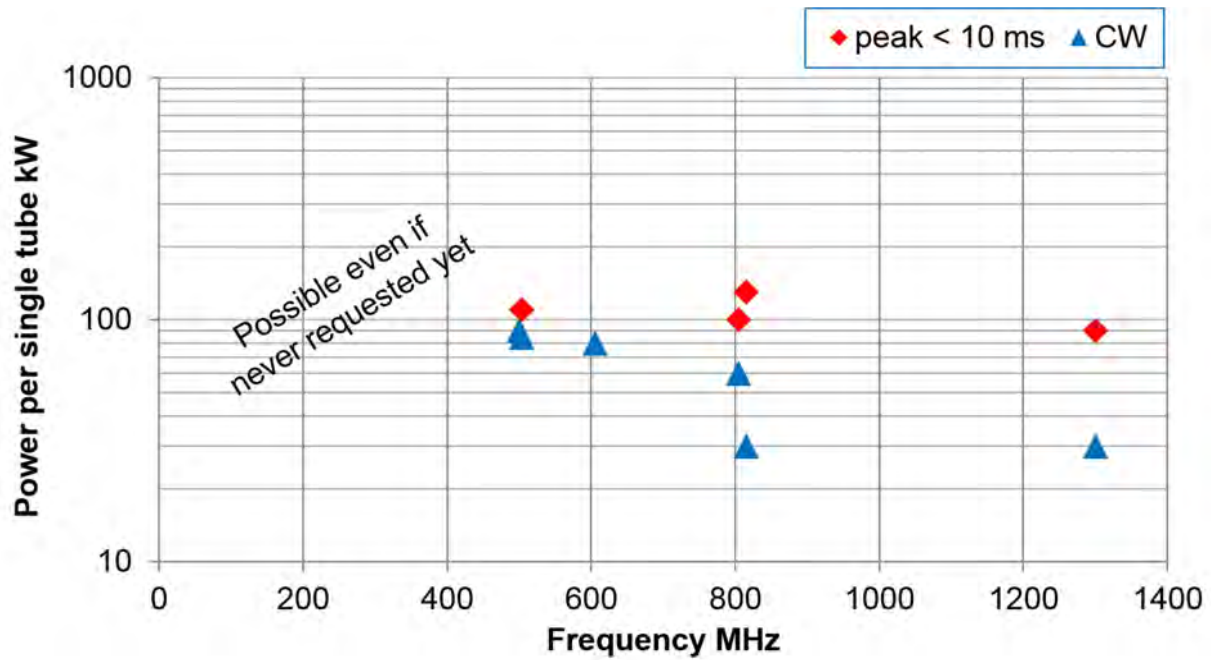

**Fig. 26:** IOTs available from industry

As we can see in Fig. 26, there are not a lot of IOTs available on the market. However, they offer a lot of possibilities in the range of few a MHz to 1.5 GHz with a power level of 20 kW to 100 kW.

## 3.3 Transistors

The last technology discussed in this paper is the SSA (Fig. 27). Although the theory was developed at almost the same time as the tubes, the construction capability came much later. Since the middle of the last century, transistors have never stopped improving. With the arrival of the mobile telephone and the digital TV broadcast, transistors have been developed in huge quantities and with increased power capabilities. They are still improving considerably, and new materials are very promising, allowing incredibly high power levels per single unit to be reached. The following list contains the main developments:

- 1925  Theory, Julius Edgar Lilienfeld [14];
- 1947  Germanium US first transistor, John Bardeen, Walter Brattain [15];
- 1948  Germanium European first transistor, Herbert Mataré and Heinrich Welker [16];
- 1953  first high-frequency transistor, Robert Wallace [17];
- 1954  Silicon transistor [18];
- 1960  Metal Oxide Semiconductor (MOS) [19];
- 1966  Gallium arsenide (GaAs) [20];
- 1980  Vertical Diffused MOS (VDMOS) [21];
- 1989  Silicon-Germanium (SiGe);
- 1997  Silicon carbide (SiC) [22];
- 2004  Carbon graphene [23].

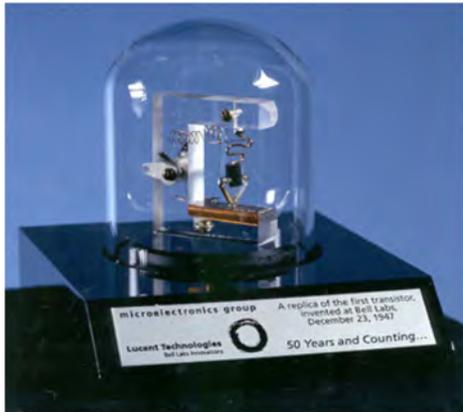
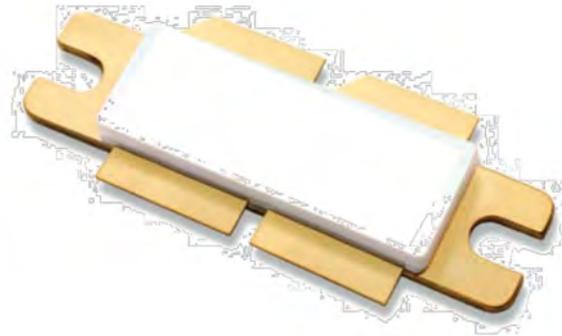

**Fig. 27:** From the first Germanium transistor in 1947 to recent LDMOS transistors in the 2000s

The conventional amplifier circuitry with transistors is the push–pull amplifier. In a push–pull circuit, the RF signal is applied to two devices. One of the devices is active on the positive voltage swing and off during the negative voltage swing. The other device works in the opposite manner so that the two devices conduct half the time. The full RF signal is then amplified. One of the main difficulties in making an RF amplifier, with this circuitry, is the use of two different types of device, one negative-positive-negative (NPN) transistor and one positive-negative-positive (PNP) transistor. Intrinsic differences in the devices will naturally introduce disturbances. Figure 28 describes such a circuit.

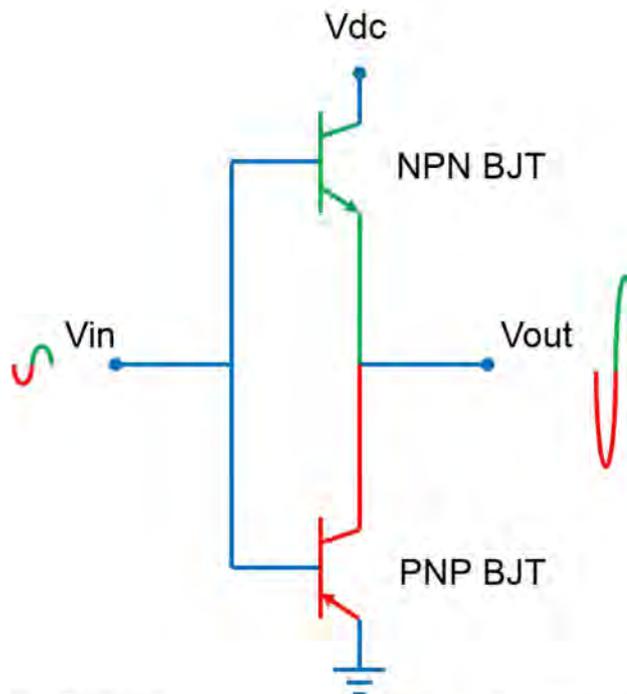

**Fig. 28:** Bipolar junction transistor (BJT) push–pull circuit

Another push–pull configuration is the balun (balanced–unbalanced) circuit (Fig. 29). Such a circuit acts as a power splitter, equally dividing the input power between the two transistors. The balun keeps one port in phase and inverts the second port in phase. As the signals are out of phase, only one device is on at a time. This configuration is easier to manufacture since only one type of device is required, and so, if the balun is correctly calculated, no disturbances are generated. This is the way most of the RF SSA are designed.

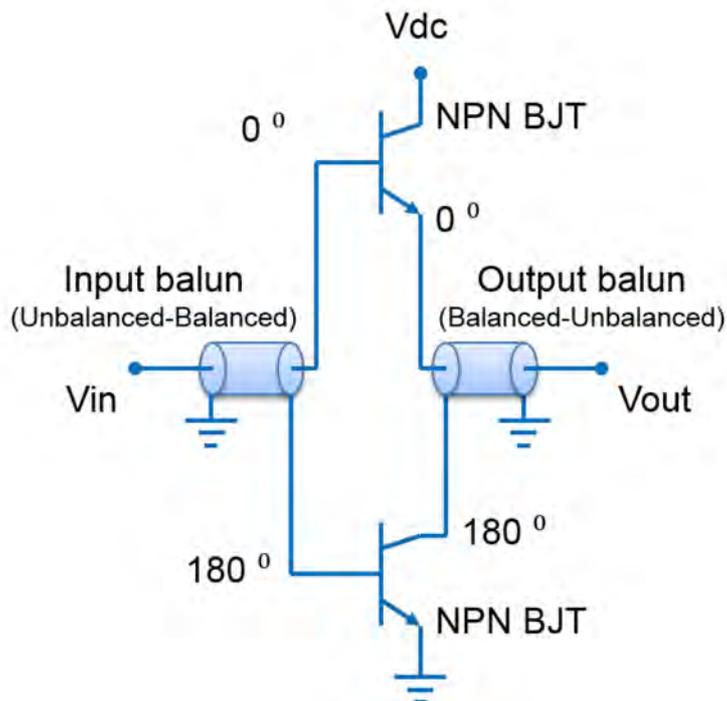

**Fig. 29:** Transistor balun circuit

The power level per unit is quite small compared to vacuum tubes. Figure 30 shows the transistors available on the market, and we can see that the frequency range starts as with the tubes and extends further compared to the tetrode, and even compared to the IOT.

If we want to build an RF high-power amplifier, we will have to combine transistors together (see Section 4). A comparison of 100 transistors with grid tubes and IOTs shows that the power level per unit is within the same power-level range. Figure 31 shows the achievable power levels from 100 combined transistors.

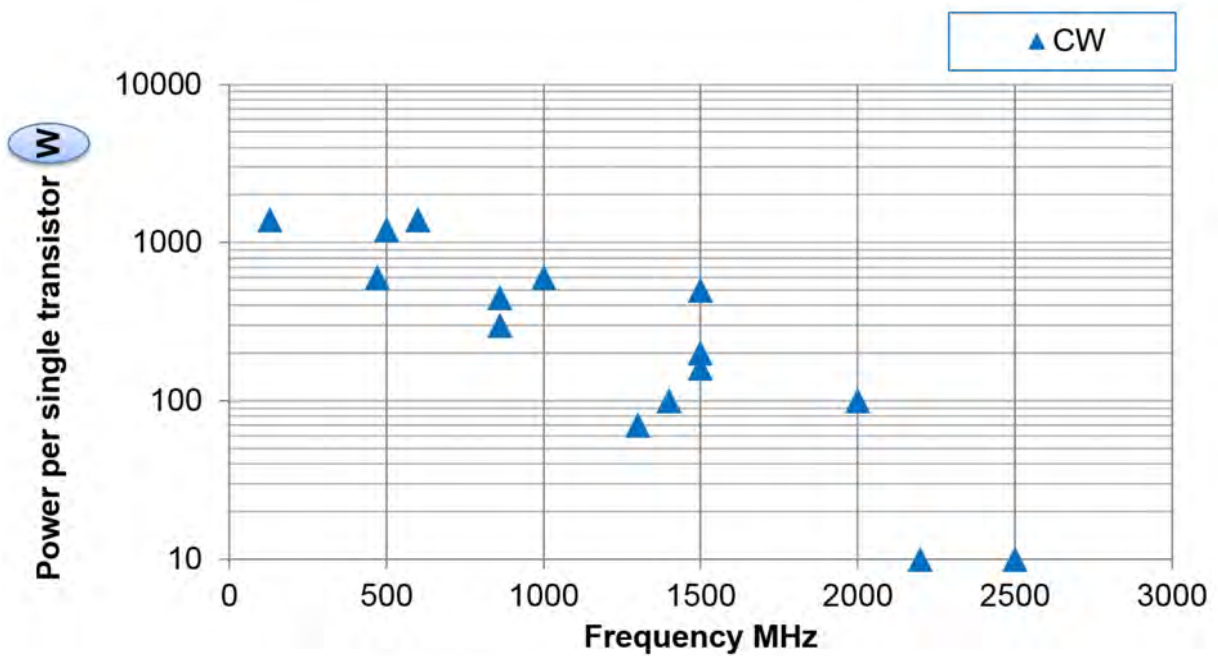

**Fig. 30:** Transistors available from industry

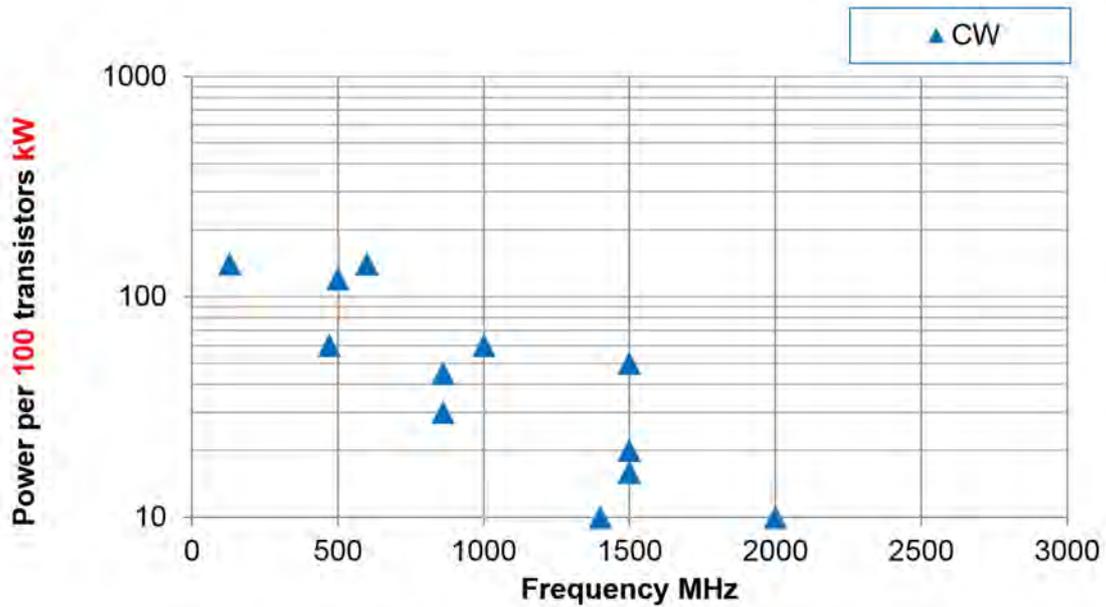

**Fig. 31:** Power level available from combining 100 transistors

## 3.4 Power overhead

Once we have defined a power system, overheads will have to be taken into consideration. Indeed, losses induced by the transmission lines, discrepancies between single units, the need for Low-Level RF (LLRF) regulations, and the fluctuation of electrical mains must be anticipated. It is commonly acknowledged that klystrons are operated at 30% (approximately –1 dB to –1.5 dB) below their maximum ratings in order to avoid running in saturation mode. Because SSPA are operated at only 10% below their maximum ratings, thanks to the granularity they offer, a fault does not affect their overall parameters so much. Tetrodes and IOTs are operated at 20% below their maximum ratings, keeping in mind that these grid tubes can be operated well over the nominal characteristic levels in pulsed mode. Figure 32 illustrates these limitations.

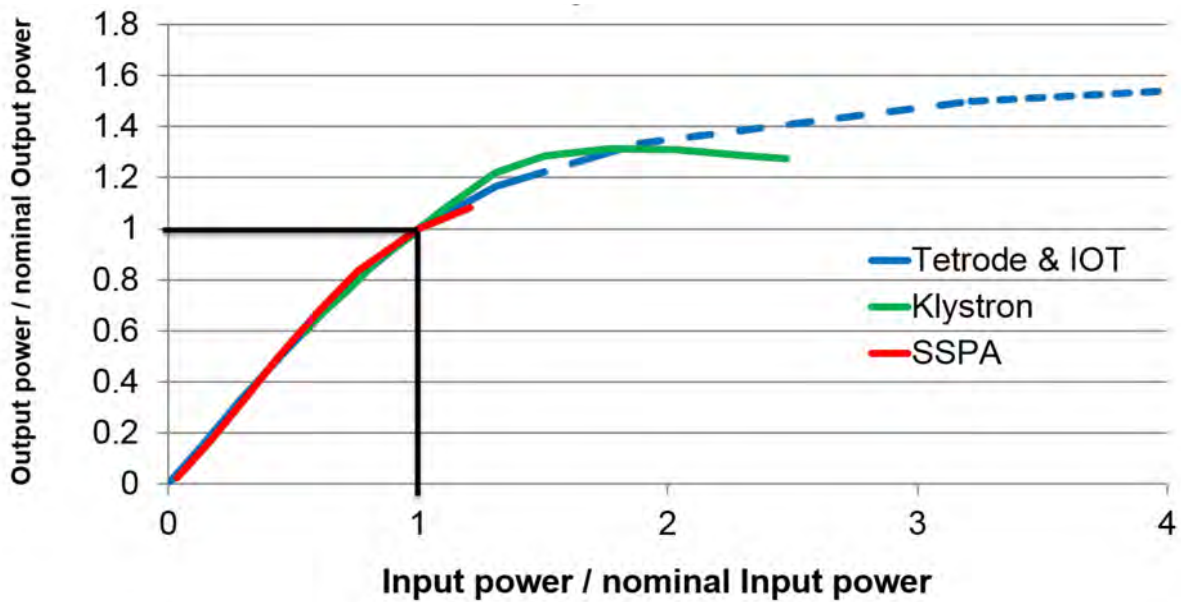

**Fig. 32:** Normalized power capability over nominal power level. SSPA cannot be operated much above their nominal ratings without being damaged. Klystrons saturate above 30 % above their nominal ratings. Grid tubes, including IOTs, allow operation above nominal ratings, even much higher in short-pulse mode.

## 3.5 Combiners and splitters

Once the RF-power amplifier source has been selected, it could be necessary to sum, or to divide, the output power of the device (Fig. 33). In RF, most of the power-combiner and power-splitter devices are reversible.

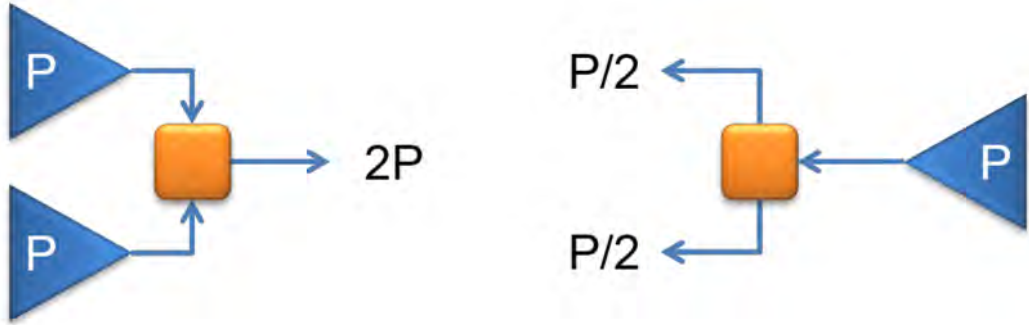

**Fig. 33:** The same device can be used as a power combiner or as a power splitter

The cheapest and easiest combiners to build are the resistive power splitters and combiners. In order to keep the correct impedance seen by all ports, they are built from resistors. Unfortunately, they are not really suited for high-power applications due to the power limitation of the resistor and to losses induced by the resistors.

The commonly used devices for power applications are the hybrid combiners. They are built from RF transmission lines and provide low losses. The power limitation of such combiners and splitters is mainly governed by the size of the lines themselves. A perfect 3 dB phase combiner (Figs. 34 and 35), with correct input phases, will allow the same power applied on each input port to be summed:

$$\Sigma = \frac{P_1 + P_2}{2} + \sqrt{P_1 P_2} \, , \tag{11}$$

$$\Delta = \frac{P_1 + P_2}{2} - \sqrt{P_1 P_2} \, . \tag{12}$$

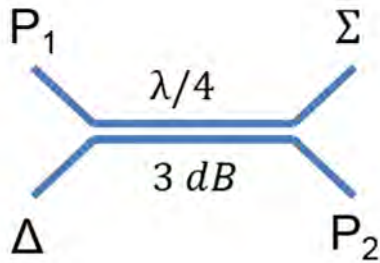

**Fig. 34:** Configuration of the 3 dB phase combiner

Correctly adjusting the phase and the gain, $P_1 = P_2 = P$:

$$\Sigma = \frac{P + P}{2} + \sqrt{PP} = 2\,P \, , \tag{13}$$

$$\Delta = \frac{P+P}{2} - \sqrt{PP} = 0. \tag{14}$$

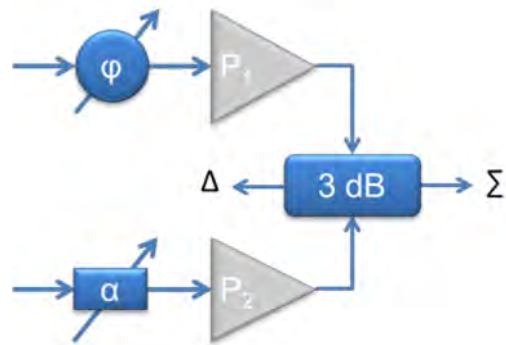

**Fig. 35:** With a phase shifter on one input line and an attenuator on the second input line, phase and gain can be adjusted to obtain a perfect 3 dB combiner.

Figure 36 shows one of the CERN SPS combiner operating at 200 MHz.

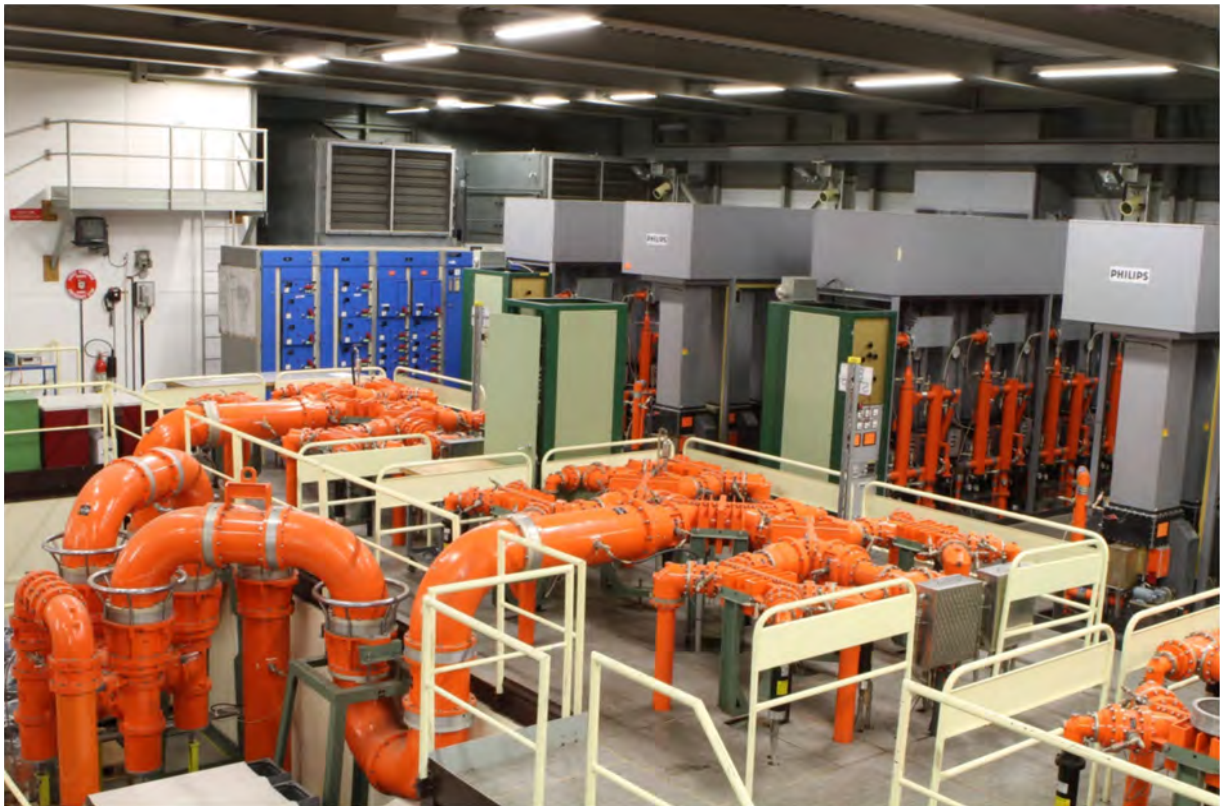

**Fig. 36:** CERN SPS 64 to 1 combiner @ 200 MHz

Another way to make a combiner (or a splitter) is the low-loss T-Junction as shown in Fig. 37. With $Z_{\lambda/4} = Z_c \sqrt{N}$, we obtain an N-ways splitter.

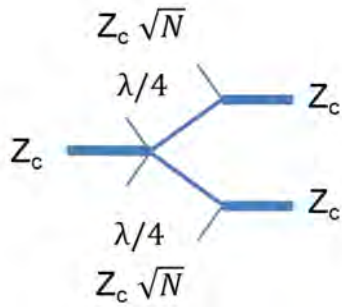

**Fig. 37:** The T-junction configuration

## 4 RF power lines

Once we have the required RF output power, it has to be transported from the RF-amplifier output to the load. Several transmission lines exist. The main lines that are used in high-power RF are the rectangular waveguides (Fig. 38) and the coaxial lines.

### 4.1 Rectangular waveguides

The main advantage of the waveguides is that waveguides provide propagation with low loss.

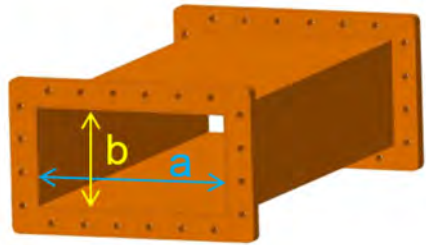

**Fig. 38:** Example of a rectangular waveguide. The size *a* is the width, and the size *b* is the height

The main parameters of a rectangular waveguide are given by the following formulas:

waveguide wavelength $$\lambda_g = \frac{\lambda}{\sqrt{1-\left(\frac{\lambda}{2a}\right)^2}},  \quad (15)$$

cut-off frequency dominant mode $$f_c = \frac{c}{2a}, \quad (16)$$

cut-off frequency next higher mode $$f_{c2} = \frac{c}{4a}, \quad (17)$$

usable frequency range $$1.3\, f_c \text{ to } 0.9\, f_{c2}. \quad (18)$$

From the transmission line, Eqs. (15)–(18), we can see that the waveguides are usable only over certain frequency ranges. For very low frequencies, the waveguide dimensions become impractically large. For very high frequencies, the waveguide dimensions become impractically small and the manufacturing tolerance becomes a significant portion of the waveguide size. Figure 39 lists some of the currently used waveguide sizes. The EIA standard names the waveguides with respect to their width, so a WR2300 waveguide has a width of 23.00 inches. This is very convenient to quickly identify the size and the reference of the waveguides. It is also common to have a half-height waveguide, when the RF power is not too high. Indeed, as it can been seen within Eqs. (15)–(17), the height is not providing any limitation.

| Waveguide name | | | Recommended frequency band of operation (GHz) | Cutoff frequency of lowest order mode (GHz) | Cutoff frequency of next mode (GHz) | Inner dimensions of waveguide opening (inch) |
| --- | --- | --- | --- | --- | --- | --- |
| EIA | RCSC | IEC | | | | |
| WR2300 | WG0.0 | R3 | 0.32 — 0.45 | 0.257 | 0.513 | 23.000 × 11.500 |
| WR1150 | WG3 | R8 | 0.63 — 0.97 | 0.513 | 1.026 | 11.500 × 5.750 |
| WR340 | WG9A | R26 | 2.20 — 3.30 | 1.736 | 3.471 | 3.400 × 1.700 |
| WR75 | WG17 | R120 | 10.00 — 15.00 | 7.869 | 15.737 | 0.750 × 0.375 |
| WR10 | WG27 | R900 | 75.00 — 110.00 | 59.015 | 118.03 | 0.100 × 0.050 |
| WR3 | WG32 | R2600 | 220.00 — 330.00 | 173.571 | 347.143 | 0.0340 × 0.0170 |

**Fig. 39:** Some of the commonly used standard waveguides

The peak-power limitation for such waveguides is given by the following formula:

$$P = 6.63 \times 10^{-4} \, E_{max}^2 \sqrt{b^2 \left( a^2 - \frac{\lambda^2}{4} \right)}, \qquad (19)$$

with

$P$ = Power in watts,

$a$ = width of waveguide in cm,

$b$ = height of waveguide in cm,

$\lambda$ = free space wavelength in cm, and

$E_{max}$ = breakdown voltage gradient of the dielectric filling the waveguide in V/cm (for dry air 30 kV/cm, for ambient air 10 kV/cm).

Looking at this formula, we can see that each waveguide size will have its own characteristics. Figure 40 illustrates the peak power limitation per waveguide size and Fig. 41 illustrates the attenuation per waveguide size.

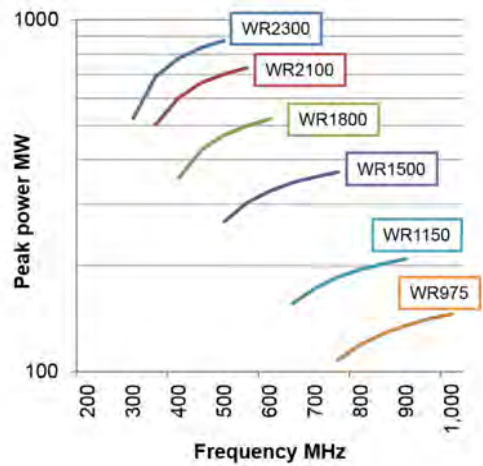

**Fig. 40:** RF frequency range and peak power per waveguide size

The attenuation of the line is dependent on the geometry, including the height, of the material:

$$\text{attenuation} = \frac{4a_0}{a} \frac{\sqrt{c/\lambda}}{\sqrt{1-(\lambda/2a)^2}} \left( \frac{a}{2b} + \frac{\lambda^2}{4a^2} \right), \qquad (20)$$

with

$a_0 = 3 \times 10^{-7}$ dB/m, for copper,

$a$ = width of waveguide in meters,

$b$ = height of waveguide in meters, and

$\lambda$ = free space wavelength in meters.

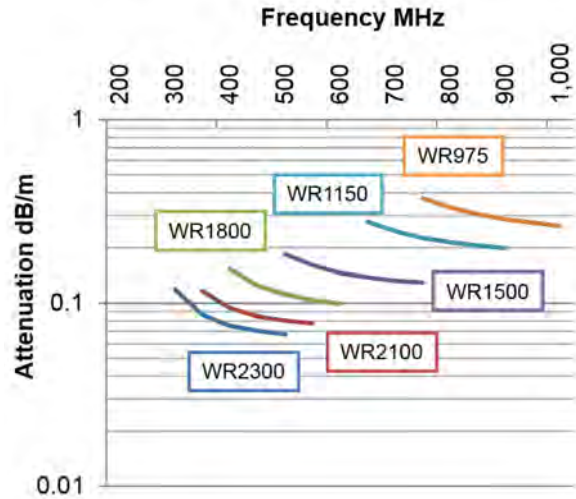

**Fig. 41:** Attenuation of a copper waveguide full height size

### 4.2 Coaxial lines

As we have seen, for lower frequencies, the waveguide dimensions become impractically large. The coaxial lines are then one of the most commonly used solutions. The characteristic impedance of a coaxial line is given by the following formula:

$$Z_c = \frac{60}{\sqrt{\varepsilon_r}} \ln\left(\frac{D}{d}\right), \qquad (21)$$

with

$D$ = inner dimension of the outer conductor,

$d$ = outer dimension of the inner conductor, and

$\varepsilon_r$ = dielectric characteristic of the medium.

The dielectric characteristic of the medium plays a very important role in a coaxial line and it is not to be neglected. Indeed, coaxial cables often have PTFE foam to keep concentricity, flexible line spacers helicoidally placed all along the line, and rigid lines made of two rigid tubes with supports to maintain concentricity. Regarding the spacers used (Fig. 42), the size of the inner and outer diameters will have to be compensated.

| Material | $\varepsilon_r$ | tan δ | Breakdown MV/m |
|---|---|---|---|
| Air | 1.00006 | 0 | 3 |
| Alumina 99.5% | 9.5 | 0.00033 | 12 |
| PTFE | 2.1 | 0.00028 | 100 |

**Fig. 42:** Some of the commonly used standard coaxial lines

Power handling of a coaxial line is related to the medium (Fig. 43) and to the breakdown field $E$. The peak-power limitation for such coaxial lines is given by the following formulas:

$$V_{\text{peakmax}} = E \frac{d}{2} \ln\left(\frac{D}{d}\right), \tag{22}$$

$$P_{\text{peakmax}} = \frac{V_{\text{peakmax}}^2}{2Z_c}, \tag{23}$$

$$P_{\text{peakmax}} = \frac{E^2 d^2 \sqrt{\varepsilon_r}}{480} \ln\left(\frac{D}{d}\right), \tag{24}$$

with

$E$ = breakdown strength of air ('dry air' $E = 3$ kV/mm, commonly used value is $E = 1$ kV/mm for ambient air),

$D$ = inside electrical diameter of outer conductor in mm,

$d$ = outside electrical diameter of inner conductor in mm,

$Z_c$ = characteristic impedance in Ω,

$\varepsilon_r$ = relative permittivity of dielectric, and

$f$ = frequency in MHz.

| Material | $\varepsilon_r$ | tan δ | Breakdown MV/m |
|---|---|---|---|
| Air | 1.00006 | 0 | 3 |
| Alumina 99.5% | 9.5 | 0.00033 | 12 |
| PTFE | 2.1 | 0.00028 | 100 |

**Fig. 43:** Peak power capability of coaxial lines strongly depends on the medium

The attenuation of a coaxial line can be approximated with the following expression:

$$\alpha = \left(\frac{36.1}{Z_c}\right)\left(\frac{1}{D} + \frac{1}{d}\right)\sqrt{f} + 9.1 \sqrt{\varepsilon_r} \, \tan\delta \, f, \tag{25}$$

where

$\alpha$ = attenuation constant, dB/m,

$Z_c$ = characteristic impedance in $\Omega$,

$f$ = frequency in MHz,

$D$ = inside electrical diameter of outer conductor in mm,

$d$ = outside electrical diameter of inner conductor in mm,

$\varepsilon_r$ = relative permittivity of dielectric, and

$\tan \delta$ = loss factor of dielectric.

When selecting the coaxial line, the best compromise has to be made with respect to the needs of the project between size, peak power capability, and attenuation. It is always very important to take all the parameters into consideration and to remember that the power limitations given by the suppliers must be carefully, and strictly, followed. Damage to a coaxial line, by mechanical deformation or by overheating, will change its impedance characteristic, and this will considerably reduce its power-handling capability.

## 4.3  Reflection from load and circulator

A major phenomenon that has to be considered when selecting the correct transmission line is the matching of the impedance. The standing wave ratio (SWR) is a measure of impedance matching of the Device under test (DUT). A wave is partly reflected when a transmission line is terminated with anything other than a pure resistance equal to its characteristic impedance (Figs. 44 and 45).

The reflection coefficient is defined by

$$\Gamma = \frac{V_r}{V_f} \ . \qquad (26)$$

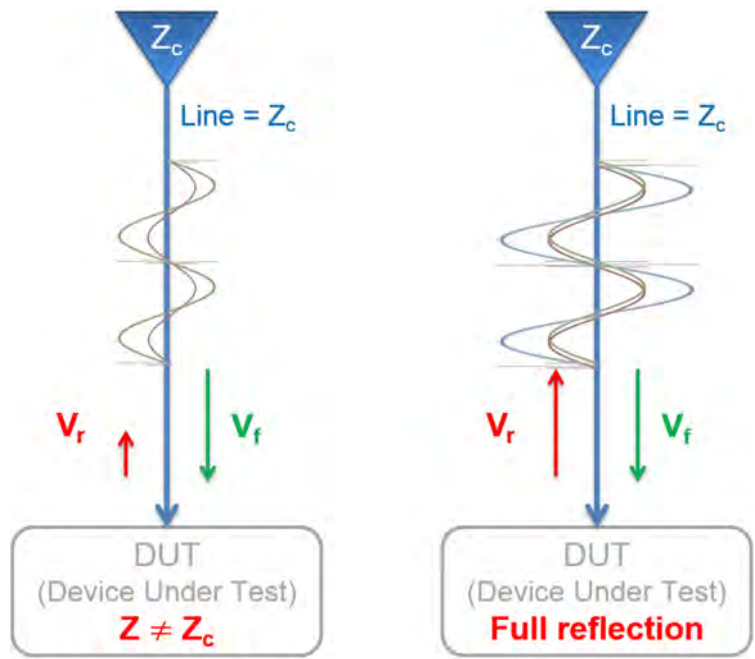

**Fig. 44:** Forward and reflected waves regarding the impedance of the DUT

| | |
|---|---|
| $\Gamma = -1$ | when the line is short-circuited complete negative reflection |
| $\Gamma = 0$ | when the line is perfectly matched, no reflection |
| $\Gamma = 1$ | when the line is open-circuited complete positive reflection |

**Fig. 45:** SWR regarding the impedance of the DUT

At some points along the line the forward and reflected waves are exactly in phase, and then

$$|V_{max}| = |V_f| + |V_r| = |V_f| + |\Gamma V_f| = (1 + |\Gamma|)\,|V_f|. \tag{27}$$

In the case of full reflection,

$$|V_{max}| = 2\,|V_f|. \tag{28}$$

At other points the forward and reflected waves are 180° out of phase, and then

$$|V_{min}| = |V_f| - |V_r| = |V_f| - |\Gamma V_f| = (1 - |\Gamma|)\,|V_f|. \tag{29}$$

In the case of full reflection,

$$|V_{min}| = 0. \tag{30}$$

The voltage standing wave ratio (VSWR) is defined by

$$\text{VSWR} = \frac{|V_{max}|}{|V_{min}|} = \frac{1 + |\Gamma|}{1 - |\Gamma|}. \tag{31}$$

So, we have seen that in case of full reflection (28), $V_{max} = 2\,V_f$. This means that $P_{max}$ is equivalent to 4 $P_f$. Here, we have to be careful, as RF-power people are often stating that the maximum power 'is' four times greater in the transmission line, in the case of full reflection. This is an abuse of language, as it is only the voltage that is doubled, and the current that is doubled, but both are not in phase under these full reflection conditions. We must remain very careful and clearly state that maximum power is equivalent to four times the forward power. An explanation is that all the datasheets from the suppliers are given in power, not in voltage. So, if we want to operate a system that has to sustain full reflection, we have to select the correct line from the supplier taking into account four times the forward power.

In any case, RF-power amplifiers will not like this reflected wave. Klystron output cavities are disturbed and grid tube, IOT, and transistor voltage capabilities have to be fully respected, so as not to damage the devices. A swift protection, if $P_r > P_{rmax}$, has to be implemented. Unfortunately, this solution, as illustrated in Fig. 46, makes the system non-operational, which is not always acceptable.

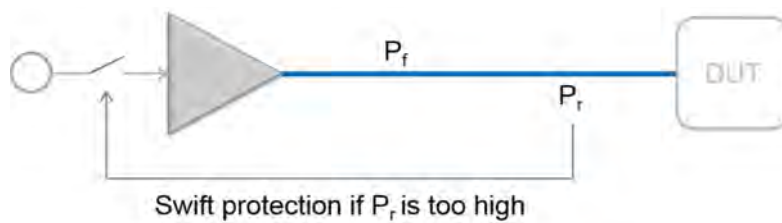

**Fig. 46:** Swift protection to protect the DUT

In order to protect our lines and our amplifiers from this reflected power, a specific component, the circulator, has been developed. This is a passive, non-reciprocal, three-port device. As shown in Fig. 47, the signal entering any port is transmitted only to the next port in rotation. The best place to insert it is close to the reflection source. The lines between the circulator and the DUT shall sustain four times $P_f$ in case of full reflection. A load of $P_f$ is needed on port 3 to absorb $P_r$.

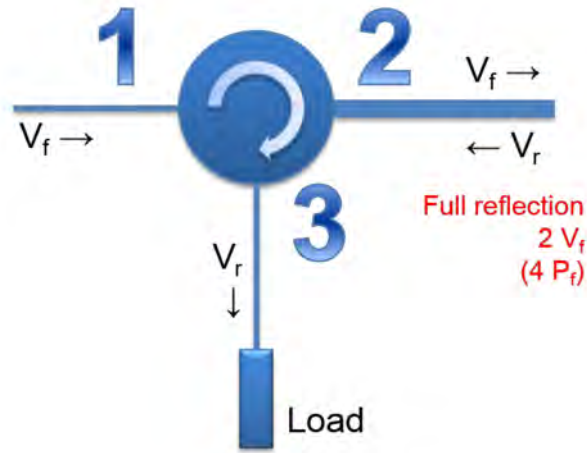

**Fig. 47:** Circulator basic principle. A device to correctly protect the DUT

Even in the case of full reflection, $V_{max} = 2\ V_f$, and so $P_{max}$ is equivalent to 4 $P_f$, the RF-power amplifiers will not see the reflected power and will not be affected. The lines between circulator and DUT must, at least, be designed for 4 $P_f$ and the load must be designed for $P_f$. The main advantage of such a configuration, shown in Fig. 48, is that the system always remains operational.

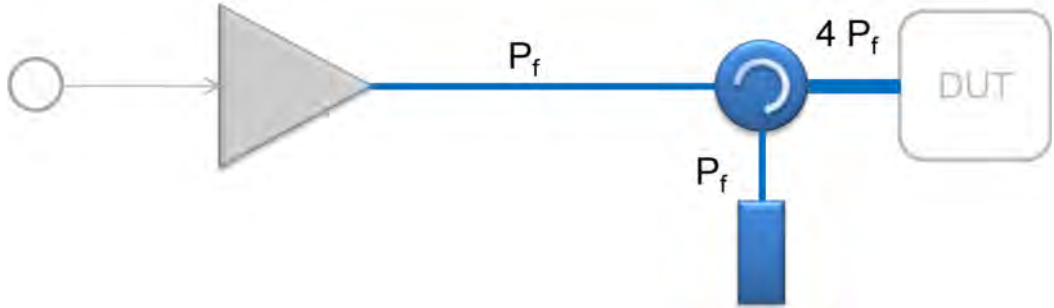

**Fig. 48:** A system protected with a circulator remains operational at any time

## 5 Fundamental Power Coupler

To complete the description of the transmission power chain, there is a final device that ensures the transfer of power from the line to the DUT: the Fundamental Power Coupler (FPC). The FPC is the connecting part between the RF transmission line and the RF cavity. It is a specific piece of transmission line that also has to provide the vacuum barrier for the beam vacuum. FPCs are one of the most critical parts of the RF-cavity system in an accelerator. A good RF design, a good mechanical design, and a high-quality fabrication are essential for efficient and reliable operation. Illustrated in Fig. 49 are some of the FPCs recently designed at CERN.

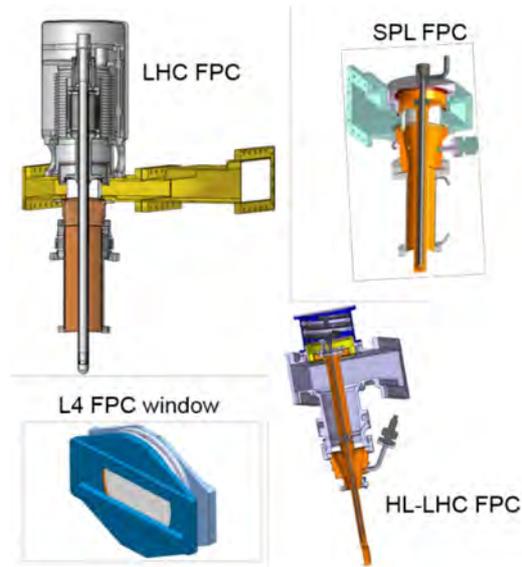

**Fig. 49:** Various CERN FPCs developed in the past few years

It is a large topic that deserves an entire document to address it. More details can be found in previous CAS listed in the References.

## 6   Case study for medical application

In order to help students to define what would be the best RF-power system, here are few illustrations to summarize the main parameters that are:

- frequency;
- overhead, peak, and average power;
- efficiency;
- rough cost estimate.

One of the first questions to address, that will define the choice of the technology, is what will be the operating frequency, or frequencies, of your machine. Figure 50 sums all the technologies available within a single plot.

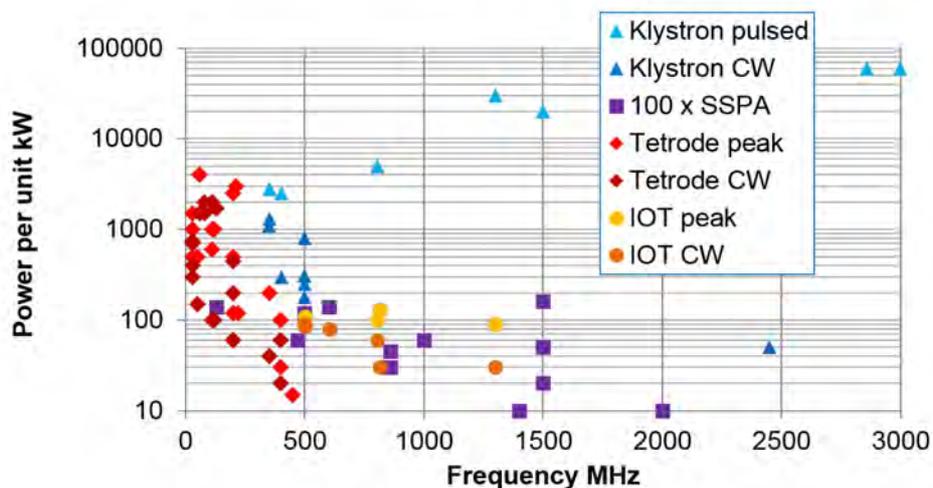

**Fig. 50:** All available RF-power sources

The second question to be answered is the overhead needed for your project, keeping in mind that one additional dB will directly add an additional 25% to the cost of an RF-power station. Peak or CW operation will also drive the selection of the RF-power source. Grid tubes, including IOTs, offer the capability of a pulsed system at much lower overhead cost, especially in short-pulse-mode operation. Figure 32 shows the overhead ratio commonly applied in the RF-power station definition for reliable operation.

Another very important issue is the overall efficiency. Usually the following numbers are applied.

- $P_{RF_{in}} \simeq 1$ to 5% $P_{RF_{out}}$ as the gain of the last amplifier stage is usually high compared to its driver stage.
- $\eta_{RF/DC} \simeq 65\%$ including overhead.
- $\eta_{PAC/PDC} \simeq 95\%$ to 98 % as power converters are of very high-quality nowadays. If the system is a pulsed system, with klystrons, this parameter has to be corrected to a lower value. Indeed, high voltage will have to be modulated and an additional fraction of the powering will be lost as the voltages must be established before the RF is applied. Figure 52 describes the phenomenon.
- Amplifier cooler $\simeq 15\%$ $P_{RF_{out}}$. This reduces considerably the overall efficiency, and it shall not be neglected.
- Building cooler $\simeq 30\%$ $P_{RF_{out}}$. This is one of the most costly parameters.

Finally the overall efficiency, also described in Fig. 51, is given by the following formula:

$$\text{overall efficiency} = \frac{P_{RF_{out}}}{P_{RF_{in}} + P_{AC_{in}} + P_{coolers}} \simeq \frac{P_{RF_{out}}}{P_{RF_{out}}(0.05 + 1.62 + 0.45)} \simeq 45\% \quad (32)$$

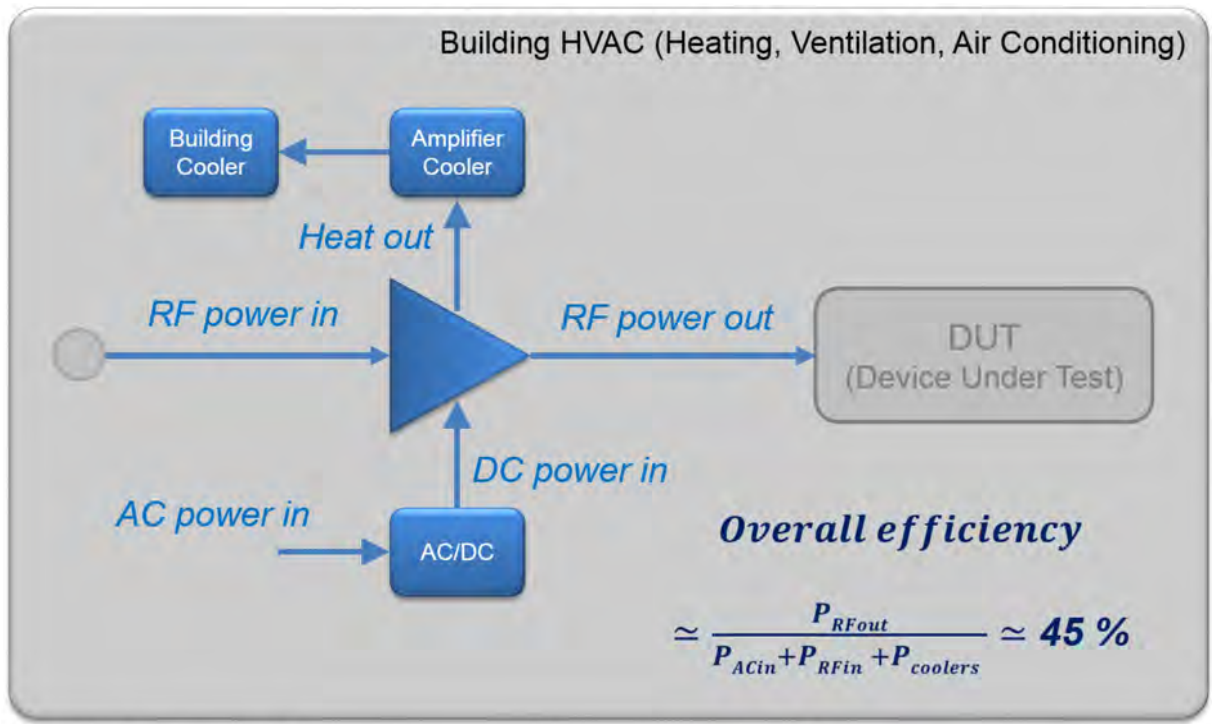

**Fig. 51:** Overall efficiency, RF power amplifier is only a fraction of the parameters to be considered

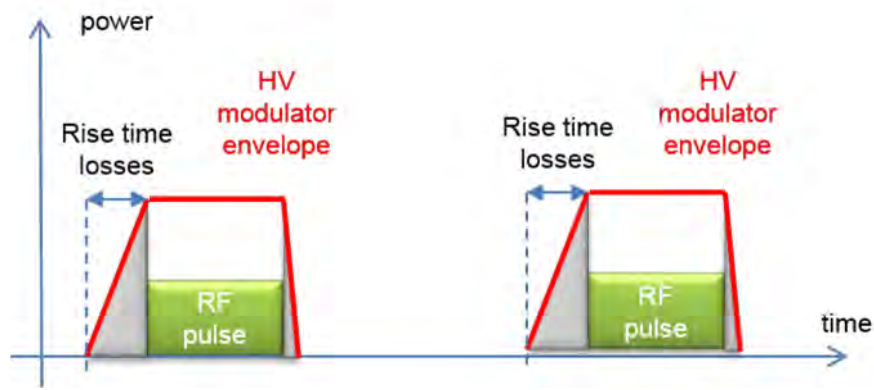

**Fig. 52:** In case of pulsed system, $\eta_{P_{AC}/P_{DC}}$ has to be corrected depending the accepted rise time and consequent loss of efficiency.

Finally, operational costs, maintenance costs, and electrical consumption costs will have to be studied in detail in order to validate the proposed RF-power system. Figure 53 provides some numbers for a given frequency and a given power level that allows all technologies to be compared.

| Technology *<br>Including SSPA driver | Very rough estimates for a 100 kW CW 352 MHz RF system<br>including RF power + Power Supplies + circulators + cooling + controls (lines not included) | Lifetime **<br>x 1000 hours | 20 years Maintenance<br>Tubes, HVPS, workshop | 20 years Electrical bill<br>3000 hours / year<br>10 hours/day<br>6/7 days<br>50 weeks/year<br>0.15 € / kWh<br>η = 45 % | Total<br>20 years |
|---|---|---|---|---|---|
| Tetrode | 500 k€ | 20 | 350 k€ | 200 k€ | 1050 k€ |
| IOT | 600 k€ | 50 | 200 k€ | 200 k€ | 1000 k€ |
| Klystron | 750 k€ | 100 | 100 k€ | 200 k€ | 1050 k€ |
| SSPA | 850 k€ | 200 | 50 k€ | 200 k€ | 1100 k€ |
| Circulator | 75 k€ | - | - | - | 75 k€ |
| Lines | 1 k€/m | - | - | - | 1 k€/m |

**Fig. 53:** Comparison of various technologies costs with a given frequency and power level. Notice that the circulator and lines are at very low cost compare to the overall costs. They must be carefully sized.

In conclusion, to design an RF-power system, you will have to carefully consider:

– your infrastructure which leads into additional overall costs;
– what power specialists are available—this will also drive your technology choice;
– the size of the transmission lines—this point is, unfortunately, often neglected, leading to deep difficulties for the whole project because of a 'simple' sub-system;
– the need, or not, for a circulator;
– your HVAC system, as this will dominate your wall–plug efficiency ratio.

## Acknowledgements

The author would like to thank Erk Jensen, CERN RF group leader, for his constant support and advice; Hans-Peter Kindermann, his CERN RF team former supervisor for his teaching; and all members of his current team for the daily achievements.